\documentclass[aps,prl,twocolumn,10pt,showpacs,superscriptaddress,amsmath,amssymb]{revtex4}
\usepackage{graphicx}
\usepackage{amsmath}  
\usepackage{amsfonts} 
\usepackage{dcolumn}
\usepackage{bm}

\begin{document}

\title{Two Poorly Measured Quantum Observables as a Complete Set of Commuting Observables}

\author{Mark Olchanyi} 
\affiliation{Newton South High School, Newton, MA 02459, USA}
\email{markolchanyi@gmail.com}

\author{Eugene Moskovets} 
\affiliation{SESI/MassTech Inc.,  Columbia, MD 21046, USA}
\email{moskovets@apmaldi.com}

\date{\today}

\begin{abstract}
In this article, we revisit the century-old question of the minimal set of observables needed to identify a quantum state: here, we replace the natural coincidences in their spectra by effective ones, induced by an imperfect measurement. We show that if the detection error is smaller than the mean level spacing, then two observables with Poisson spectra will suffice, no matter how large the system is. The primary target of our findings is the integrable (i.e. exactly solvable) quantum systems whose spectra do obey the Poisson statistics. We also consider the implications of our findings for classical pattern recognition techniques.
\end{abstract}

\maketitle

\section{Introduction}
In quantum physics, the state of the system
can be unambiguously determined by measuring several select integrals of motion (observables conserved in time evolution);
such observables are said to form
a complete set of commuting observables (CSCO) \cite{gasiorowicz1974}.
Mathematically, the
spectra of the allowed values of the members of the set $\left\{\hat{I}^{(1)},\, \hat{I}^{(2)},\,\ldots,\,\hat{I}^{(k)},\ldots \right\}$ are the
discrete sequences of real
numbers, $\left\{
	\{I^{(1)}_{1},\,I^{(1)}_{2},\,\ldots\},\,
	\{I^{(2)}_{1},\,I^{(2)}_{2},\,\ldots\},\,
	\{I^{(k)}_{1},\,I^{(k)}_{2},\,\ldots\},\,
	\ldots
	     \right\}$,
that obey the following
property: for any pair of indices $(n_{1},\,n_{2})$, there exists
at least one member of the CSCO (say $\hat{I}^{(k)}$) for which the corresponding elements are non-degenerate
(i.e. distinct, $I^{(k)}_{n_{1}} \neq I^{(k)}_{n_{2}})$.
Operationally, for any $\tilde{n}$, the knowledge of the real-number sequence
$\left\{ I^{(1)}_{\tilde{n}},\, I^{(2)}_{\tilde{n}},\ldots,\,I^{(k)}_{\tilde{n}},\,\ldots \right\}$
is sufficient to infer what $\tilde{n}$ was.

\begin{figure}
\centering
\includegraphics[scale=.4]{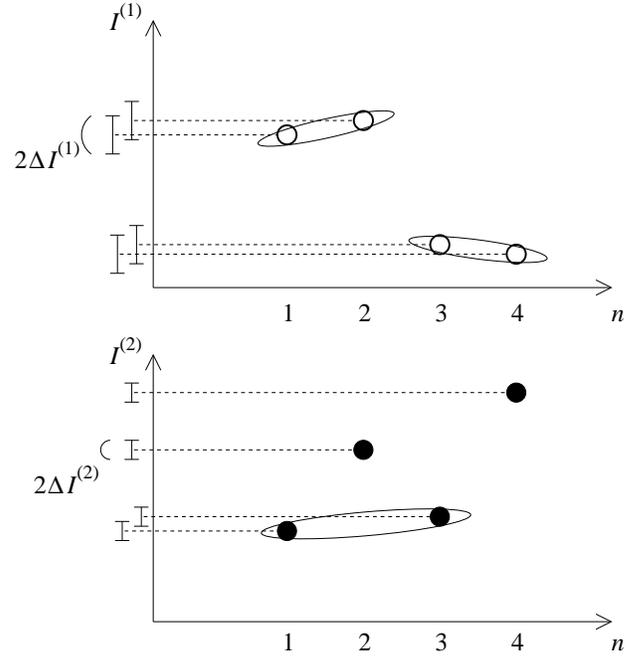}
\caption{
An example of two observables, $\hat{I}^{(1)}$ and $\hat{I}^{(2)}$,
forming a complete set of commuting observables (CSCO). Even if a particular pair
of measured values of one of the observables is degenerate (points in ovals),
i.e. indistinct given the measurement error,
it will not be degenerate vis-a-vis another observable.
If both observables are measured with
respective errors $\Delta I^{(1)}$ and $\Delta I^{(2)}$,
the state of the system, $n$, can be determined unambiguously.
        }
\label{f:CSCO_novel_ON}        
\end{figure}

\begin{figure}
\centering
\includegraphics[scale=.4]{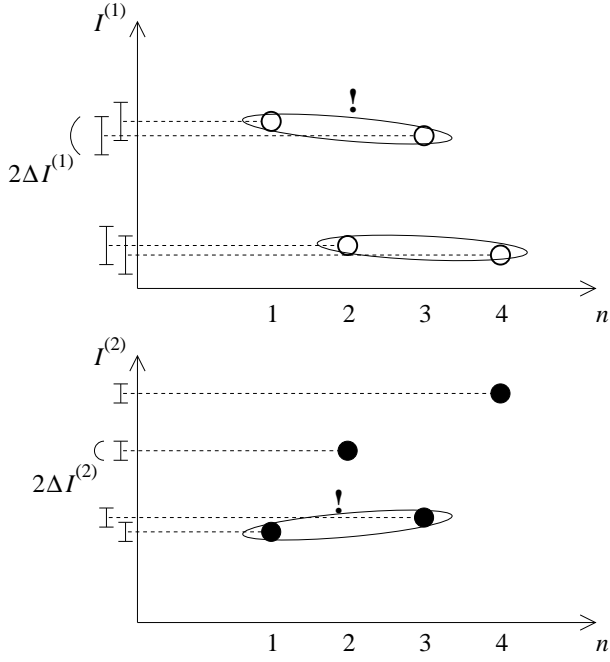}
\caption{
Two observables that do not form a CSCO.
If $\hat{I}^{(1)}$ and $\hat{I}^{(2)}$ are the
only two observables that are measured, the states $n=1$ and $n=3$ are indistinct.
        }
\label{f:CSCO_novel_OFF}
\end{figure}

In this paper we consider a situation where the indistinguishability occurs due to an insufficient accuracy of the detection.
We assume that the spectra of the CSCO members are drawn from random processes, later chosen to be of the Poisson type.
We further assume that for a given observable
$\hat{I}$, two of its possible values, $I_{n_{1}}$ and $I_{n_{2}}$, can {\it not} be resolved if they are separated by a distance less
than the detection
error, $|I_{n_{1}}-I_{n_{2}}|<\Delta I$; in this case, they are considered to be degenerate for all practical
purposes. The principal goal of this paper is
to determine the probability for two observables, $\hat{I}^{(1)}$ and $\hat{I}^{(2)}$,
to form an CSCO (see examples depicted at Figs. \ref{f:CSCO_novel_ON} and \ref{f:CSCO_novel_OFF}).


Furthermore, we will concentrate on quantum systems with finite spectra of size $N$, i.e.
we will assume that a given system can be in any of $N$ available states, where $N$ is finite. In
real-world applications, $N$ grows with the system size.

This study is partially inspired by a related issue of the functional dependence
between the quantum observables (i.e. the ability to predict the value of one observable
having had measured the values of several others):
while being a clear and extremely useful
concept in classical physics, the practical significance of the notion of functional dependence in the
quantum world is questionable at best. Already in 1929, von Neumann showed \cite{neuman2010_201} that
for any quantum system with $N$ states,
one can construct a set of as many as $N-1$ conserved quantities each functionally independent from the
other $N-2$; this is provided that  any number of degeneracies is allowed. On the other hand,
Sutherland argued \cite{sutherland2004_book} that
any conserved quantity is functionally dependent on any other conserved quantity with a non-degenerate spectrum. Being able
to determine the state of the system amounts to being able to predict the value of any other observable. Thus,
from the functional dependence perspective we are looking for the probability that all conserved quantities
be functionally dependent on two other, chosen beforehand.

\section{Statement of problem}
Consider two $N$-element-long finite (ordered) sequences of real numbers,
\begin{align*}
& \left( I^{(1)}_{1},\,I^{(1)}_{2},\,\ldots,\,I^{(1)}_{n},\,\ldots,\,I^{(1)}_{N} \right)
\\
& \left( I^{(2)}_{1},\,I^{(2)}_{2},\,\ldots,\,I^{(2)}_{n},\,\ldots,\,I^{(2)}_{N} \right)
\,\,.
\end{align*}
Both sequences are assumed to be $N$-event-long randomly reshuffled fragments of a Poisson process.
Let us elaborate on what this assumption means. Consider monotonically increasing permutations of the above sequences,
\begin{align*}
& \left( \tilde{I}^{(1)}_{1},\,\tilde{I}^{(1)}_{2},\,\ldots,\,\tilde{I}^{(1)}_{n},\,\ldots,\,\tilde{I}^{(1)}_{N} \right)
\\
&\qquad \tilde{I}^{(1)}_{n+1} > \tilde{I}^{(1)}_{n} \mbox{ for any } n
\\
& \left( \tilde{I}^{(2)}_{1},\,\tilde{I}^{(2)}_{2},\,\ldots,\,\tilde{I}^{(2)}_{n},\,\ldots,\,\tilde{I}^{(2)}_{N} \right)
\\
&\qquad \tilde{I}^{(2)}_{n+1} > \tilde{I}^{(2)}_{n} \mbox{ for any } n
\,\,.
\end{align*}
According to the definition of the Poisson process, the probability of finding exactly $k$ elements of the $\tilde{I}^{(1)}$
sequence in an interval of a length ${\cal I}$ is
\begin{align*}
&
\mbox{{\it Prob}}[\mbox{$k$ elements of $\tilde{I}^{(1)}$ in $[I,\,I+{\cal I}]$}] =
\\
&
\qquad\qquad
\frac{
	     e^{
	     	  -{\cal I}
	     	  /
	     	  \overline{\delta I^{(1)}}
	       }
	      (
	      {\cal I}
	      /
	      \overline{\delta I^{(1)}}
	      )^k
	 }
	 {
	     k!
	 }
\,\,,
\end{align*}
for any initial position of the interval $I$. Here, $\overline{\delta I^{(1)}}$ is the expectation value
of the interval between any two consecutive elements of the sequence:
\begin{align*}
\overline{\delta I^{(1)}} \equiv \mbox{{\it Mean}}[\tilde{I}^{(1)}_{n+1}-\tilde{I}^{(1)}_{n}]
\,\,.
\end{align*}
Likewise,
\begin{align*}
&
\mbox{{\it Prob}}[\mbox{$k$ elements of $\tilde{I}^{(2)}$ in $[I,\,I+{\cal I}]$}] =
\\
&
\qquad\qquad
\frac{
	     e^{
	     	  -{\cal I}
	     	  /
	     	  \overline{\delta I^{(2)}}
	       }
	      (
	      {\cal I}
	      /
	      \overline{\delta I^{(2)}}
	      )^k
	 }
	 {
	     k!
	 }
\\
&
\overline{\delta I^{(2)}} \equiv \mbox{{\it Mean}}[\tilde{I}^{(2)}_{n+1}-\tilde{I}^{(2)}_{n}]  	
\,\,.
\end{align*}

An important particular property of the Poisson processes
is that the intervals between two successive elements of the sequence are mutually statistically
independent, and they are distributed according to the {\it exponential} law:
\begin{align}
\begin{aligned}
&
\mbox{{\it Prob}}[\tilde{I}^{(1)}_{n+1}-\tilde{I}^{(1)}_{n} > {\cal I}] = e^{-{\cal I}/\overline{\delta I^{(1)}}}
\\
&
\mbox{{\it Prob}}[\tilde{I}^{(2)}_{n+1}-\tilde{I}^{(2)}_{n} > {\cal I}] = e^{-{\cal I}/\overline{\delta I^{(2)}}}
\label{exponential_distribution}
\end{aligned}
\,\,.
\end{align}

Let us now return to the original sequences $I^{(1)}$ and $I^{(2)}$ that are random permutations of
the $\tilde{I}^{(1)}$ and $\tilde{I}^{(2)}$:
\begin{align*}
&
\left( I^{(1)}_{1},\,I^{(1)}_{2},\,\ldots,\,I^{(1)}_{n},\,\ldots,\,I^{(1)}_{N} \right) =
\\
&
\quad \mbox{Random permutation}[
      \left( \tilde{I}^{(1)}_{1},\,\tilde{I}^{(1)}_{2},\,\ldots,\,\tilde{I}^{(1)}_{n},\,\ldots,\,\tilde{I}^{(1)}_{N} \right)
                                ]
\\
&
\left( I^{(2)}_{1},\,I^{(2)}_{2},\,\ldots,\,I^{(2)}_{n},\,\ldots,\,I^{(2)}_{N} \right) =
\\
&
\quad \mbox{Random permutation}[
      \left( \tilde{I}^{(2)}_{1},\,\tilde{I}^{(2)}_{2},\,\ldots,\,\tilde{I}^{(2)}_{n},\,\ldots,\,\tilde{I}^{(2)}_{N} \right)
                                ]
\,\,.
\end{align*}
It also follows from the properties of the Poisson processes that the $I^{(1)}$ and $I^{(2)}$) sequences can be
approximately generated by dropping $N$ uniformly distributed random numbers on respective intervals of lengths
$(N+1) \overline{\delta I^{(1)}}$ and $(N+1) \overline{\delta I^{(2)}}$.

For a given sequence, say $I^{(1)}$, call a pair of elements, $I^{(1)}_{n_{1}}$ and $I^{(1)}_{n_{2}}$, {\it degenerate} if
the difference between them is less than a given number, $\Delta I^{(1)}$, called the {\it detecting error}:
\begin{align*}
&
|I^{(1)}_{n_{1}} - I^{(1)}_{n_{2}}| < \Delta I^{(1)}
\,\,.
\end{align*}
Assume that the sequence $\left( I^{(1)}_{1},\,I^{(1)}_{2},\,\ldots,\,I^{(1)}_{n},\,\ldots,\,I^{(1)}_{N} \right)$ is known,
and a device (emulating a process of quantum measurement) produces a particular element of it, $I^{(1)}_{n}$. An observer
is allowed to measure it in an attempt to determine what the index $n$ was (the analogue of a quantum state index),
using the sequence as a look-up table.
For a perfect measurement device and
in the absence of coinciding elements in the sequence, this task can be easily accomplished. If however the accuracy
of the measurement is limited, and the values separated by less than $\Delta I^{(1)}$ can not be distinguished, the
value $I^{(1)}_{n}$ may happen to belong to one of the degenerate pairs, $(I^{(1)}_{n_{1}},\,I^{(1)}_{n_{2}})$.
If, further, there is no other degenerate pair $I^{(1)}_{n}$ belongs to, $n$ can only be said to be equal to {\it either}
$n_{1}$ {\it or} $n_{2}$. If there are other relevant degenerate pairs, the uncertainty can be even greater.

Likewise, degenerate pairs of elements of the second sequence is determined through
\begin{align*}
&
|I^{(2)}_{n_{1}} - I^{(2)}_{n_{2}}| < \Delta I^{(2)}
\,\,,
\end{align*}
where $\Delta I^{(2)}$ is the error of the detector that measures the $\hat{I}^{(2)}$ observable. Now, both
$I^{(1)}_{n}$ and $I^{(2)}_{n}$ are produced, and both detectors are used.

When both measurements are
allowed, several ambiguities in determining the index can be removed. However, if for a given pair of indices,
$n_{1}$ and $n_{2}$, both $(I^{(1)}_{n_{1}},\,I^{(1)}_{n_{2}})$ and $(I^{(2)}_{n_{1}},\,I^{(2)}_{n_{2}})$ constitute
the respective degenerate pairs, the
indices $n_{1}$ and $n_{2}$ can never be resolved. The goal of this work is to determine the probability
that no above ambiguities exist:
\begin{align*}
&
\mbox{{\it Prob}}[\mbox{no $(n_{1},\,n_{2})$ such that}
\\
&
\mbox{
$|I^{(1)}_{n_{1}} - I^{(1)}_{n_{2}}| < \Delta I^{(1)}$ and
$|I^{(2)}_{n_{1}} - I^{(2)}_{n_{2}}| < \Delta I^{(2)}$
} ] = ?
\,\,.
\end{align*}
Physically, we are interested in computing the probability that the observables
$\hat{I}^{(1)}$ and $\hat{I}^{(2)}$ constitute a complete set of commuting observables (CSCO).

\section{A preliminary study of isolated sequences}

\subsection{Monotonically increasing sequence}
Consider a length-$N$ fragment of a Poisson process,
\begin{align*}
& \left( \tilde{I}_{1},\,\tilde{I}_{2},\,\ldots,\,\tilde{I}_{n},\,\ldots,\,\tilde{I}_{N} \right)
\\
&\qquad \tilde{I}_{n+1} > \tilde{I}_{n} \mbox{ for any } n
\,\,,
\end{align*}
with the mean spacing $\overline{\delta I}$. In many respects, studying monotonically increasing
sequences is easier than their random-order counterparts. This is fortunate since several important conclusions about the
latter can be drawn using the former.

\subsubsection{Numerical engine}
As a model for a length-$N$ fragment of a Poisson process with the mean spacing $\overline{\delta I}$ we
use $N$ pseudorandom numbers uniformly distributed on an interval between 0 and $(N+1) \overline{\delta I}$. When
ordered in the order it has been generated, the sequence serves as a model for a random permutation
of a Poisson sequence, $\left( I_{1},\,I_{2},\,\ldots,\,I_{n},\,\ldots,\,I_{N} \right)$, the primary object of interest.
When rearranged to a monotonically increasing sequence, a model for a Poisson sequence per se,
$\left( \tilde{I}_{1},\,\tilde{I}_{2},\,\ldots,\,\tilde{I}_{n},\,\ldots,\,\tilde{I}_{N} \right)$, is generated.

\begin{figure}[!h]
\centering
\includegraphics[scale=.65]{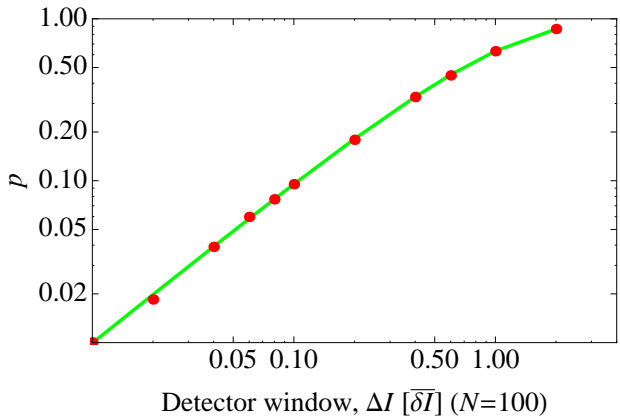}
\caption{
Probability for a given pair of consecutive elements of a Poisson-distributed spectrum to
be unresolvable by a detector with an error $\Delta I$ (i.e. to be degenerate).
$\overline{\delta I}$ is the mean spacing between consecutive elements. Red dots: numerical model,
uncorrelated, uniformly-distributed random numbers on an interval, subsequently rearranged in
a monotonically increasing succession. Green line: theoretical prediction (\ref{p}). Numerically,
the spectrum consists of 100 elements; the theoretical prediction is universal. Numerical points
result from averaging over 1000 Monte Carlo realizations.    	
        }
\label{f:_R__ProbabilityForAPairToBeDegenerateI}
\end{figure}

To test the numerical method, we compare, in Fig.~\ref{f:_R__ProbabilityForAPairToBeDegenerateI}, numerical
and analytical predictions for the probability $p$ of two subsequent elements of the monotonically increasing sequence
to form a degenerate pair, with respect to a detection error $\Delta I$. The analytic prediction,
\begin{align}
\begin{aligned}
&
p \equiv \mbox{{\it Prob}}[\mbox{for a given $n$, }\tilde{I}_{n+1}-\tilde{I}_{n} < \Delta I] =
\\
&\qquad\qquad\qquad\qquad
1-e^{-\Delta I/\overline{\delta I}}
\stackrel{\Delta I \ll \overline{\delta I}}{\approx} \frac{\Delta I}{\overline{\delta I}}
\,\,,
\label{p}
\end{aligned}
\end{align}
follows directly from one of the properties of the Poisson sequences (\ref{exponential_distribution}). The numerical
and analytical predictions agree very well.

From (\ref{p}) it also follows that the average number of degenerate pairs is
\begin{align}
\bar{M} = p N_{p} \stackrel{N\to\infty}{\approx} p N
\,\,,
\label{M_bar}
\end{align}
where 
\begin{align}
N_{p} \equiv N-1
\label{N_p}
\end{align}
is the number of pairs of indices with two neighboring elements.

\subsubsection{Clusters of degenerate pairs}
\begin{figure}
\centering
\includegraphics[scale=.65]{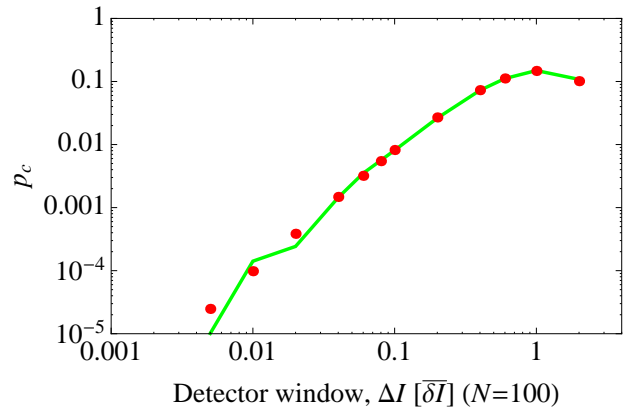}
\caption{
Probability for a given pair of consecutive elements of a Poisson-distributed spectrum to be
degenerate and to be immediately followed on the right by more degenerate pairs, as a function
of the detector error $\Delta I$.  Red dots: numerics.
Green line: theoretical prediction (\ref{p_c}). It does not depend on the sequence length $N$.	
The rest of the data is the same as in Fig.~\ref{f:_R__ProbabilityForAPairToBeDegenerateI}.
        }
\label{f:_R__ProbabilityForAPairToBeLeftPairOfDegenerateClusterI}
\end{figure}

In what follows, the analytical predictions will be greatly simplified thanks to our ability to assume that
degenerate pairs do not share elements, i.e. that a given element can either be isolated or belong to
only one degenerate pair. It is clear that the probability of forming a degenerate cluster must be much smaller than the one for
an isolated pair, because of the small probability of forming the latter. Yet, a quantitative assessment is due. Assuming that
that the individual spacings between the subsequent values of $I$ are statistically independent, we get:
\begin{align}
\begin{aligned}
&
p_{c} \equiv
\\
&\qquad
\mbox{{\it Prob}}[\mbox{for a given $n$, }(\tilde{I}_{n}-\tilde{I}_{n-1} > \Delta I)\mbox{ and }
\\
&\qquad\qquad\qquad
(\tilde{I}_{n+1}-\tilde{I}_{n} < \Delta I)\mbox{ and }
\\
&\qquad\qquad\qquad\qquad
(\tilde{I}_{n+2}-\tilde{I}_{n+1} < \Delta I)] = (1-p)p^2
\,\,.
\label{p_c}
\end{aligned}
\end{align}
Note that the above probability is the probability for the index $n$ to be at the left end of a cluster of degenerate pairs
containing more than one pair. Fig.~\ref{f:_R__ProbabilityForAPairToBeLeftPairOfDegenerateClusterI} compares the above
prediction with {\it ab initio} numerics: the agreement is remarkable. The overall conclusion is that for a small
probability $p$ of forming a degenerate pair, one can safely assume that all the degenerate pairs are isolated.
This result will be heavily used in a subsequent treatment of the more physical random-order sequences.

It is instructive to also compute the probability of having no degenerate clusters at all, for two reasons.
(a) this is the first measure in this study that characterizes the
$\left( \tilde{I}_{1},\,\tilde{I}_{2},\,\ldots,\,\tilde{I}_{n},\,\ldots,\,\tilde{I}_{N} \right)$ as a set,
with no reference to the order of the elements. The probability of no clusters we will obtain
will therefore be identical to the one for the $\left( I_{1},\,I_{2},\,\ldots,\,I_{n},\,\ldots,\,I_{N} \right)$ sequence,
randomly ordered. On the other hand, the monotonically increasing sequences are conceptually simpler, and they allow for
more intuition in designing analytical predictions. (b) Using this study we will show that while the spacings
between neighboring members of the monotonic sequence are statistically independent, the degenerate pairs they produce
are not statistically independent from the point of view of the randomly permuted sequence.

Even though the spacings between the consecutive elements of the $\tilde{I}$ sequence are statistically
independent, the appearance of a cluster with the leftmost edge at an index $n$ will be preventing an appearance of another
cluster at $n+1$. Thus, appearances of clusters at different points are not statistically independent. Nevertheless,
for a very low cluster probability, a typical distance between the clusters becomes much larger than their typical
length, and the above correlations can be neglected. Our estimate for the probability of having
no clusters at all then reads:
\begin{align}
\begin{aligned}
&
P_{\mbox{\scriptsize no clusters}} \equiv
\\
&\qquad
\mbox{{\it Prob}}[
\mbox{for all $n$, }
(\tilde{I}_{n+1}-\tilde{I}_{n} > \Delta I)\mbox{ or }
\\
&\qquad\qquad
\left(
(\tilde{I}_{n+1}-\tilde{I}_{n} < \Delta I)\mbox{ and }
\right.
\\
&\qquad\qquad\qquad
(\tilde{I}_{n}-\tilde{I}_{n-1} > \Delta I) \mbox{ and }
\\
&\qquad\qquad\qquad\qquad
\left.
(\tilde{I}_{n+2}-\tilde{I}_{n+1} > \Delta I)
\right)] \approx
\\
&\qquad (1-p_{c})^{N_{p}} \stackrel{p\to 0,\, N\to\infty,\, p^2 N \to \mbox{const}}{\approx} e^{-p_{c} N} \stackrel{p\to0}{\approx} e^{-p^2 N}
\,\,,
\label{P_no_clusters_correct}
\end{aligned}
\end{align}
where $N_{p}$ is given by (\ref{N_p}).
Above, we neglected limitations on the cluster length closer to the right edge of the spectrum, since we are mainly interested
in the $N\to\infty$ limit.

\begin{figure}
\centering
\includegraphics[scale=.65]{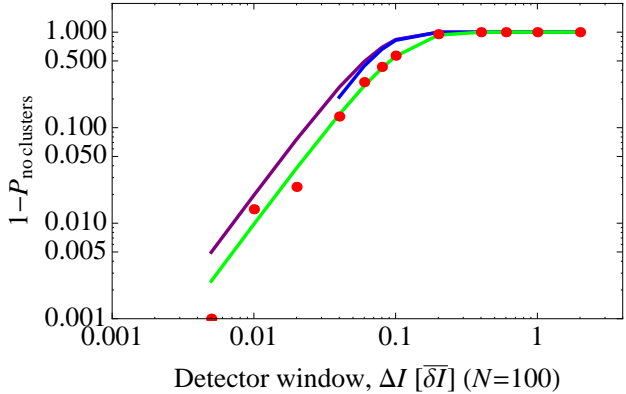}
\caption{
The probability of having no clusters of consecutive degenerate pairs of elements at all, as a function
of the detector error $\Delta I$.
Red dots: numerics.
Green line: theoretical prediction (\ref{P_no_clusters_correct}). Blue line: a naive
combinatorial hypothesis
(\ref{P_no_clusters_incorrect}). At small detector window widths, the formula involves
factorials of large numbers and becomes numerically unreliable. Purple line: the same as the blue line
but using an asymptotic expression.  The rest of
the data is the same as in Fig.~\ref{f:_R__ProbabilityForAPairToBeDegenerateI}.
        }
\label{f:_R__ProbabilityOfNoClustersI}
\end{figure}

As we have mentioned above, the result (\ref{P_no_clusters_correct}) applies for both $\tilde{I}$ and $I$ sequences, being
a characteristics of these sequences as a set, not as sequences with a particular order. It is tempting to
reinterpret it as the probability that $\bar{M}$ pairs chosen uniformly at random from
\begin{align}
{\cal N}_{p}\equiv \frac{N(N-1)}{2}
\label{cal_N_p}
\end{align}
pairs of elements of an $N$-member-strong set have no elements in common. The number of ways to choose
$\bar{M}$ pairs with no elements in common is
\begin{align*}
\frac{2^{\bar{M}} N!}{\bar{M}! (N-2\bar{M})!}
\,\,.
\end{align*}
The number of ways to chose any set of $\bar{M}$ pairs is
\begin{align*}
\frac{{\cal N}_{p}!}{\bar{M}!({\cal N}_{p}-\bar{M})!}
\,\,.
\end{align*}
Dividing the former by the latter
we get the following expression:
\begin{align}
P_{\mbox{\scriptsize no clusters, conjecture}} =
\frac{2^{\bar{M}} N! ({\cal N}_{p}-\bar{M})!}{{\cal N}_{p}! (N-2\bar{M})!}
\label{P_no_clusters_incorrect}
\\
\quad\stackrel{p\to 0,\, N\to\infty,\, p^2 N \to \mbox{const}}{\approx}
e^{-2 p^2 N}
\,\,,
\label{P_no_clusters_incorrect_asymptotic}
\end{align}
where $\bar{M}$ is given by (\ref{M_bar}).
Comparing (\ref{P_no_clusters_incorrect}) to (\ref{P_no_clusters_correct}) one notices
that the former overestimates the probability of cluster formation by a factor of two.
An interpretation of this simple relationship requires future research. Figure~\ref{f:_R__ProbabilityOfNoClustersI} illustrates this difference.
\subsection{Implications for a randomly permuted counterpart}

The principal goal of our paper is to assess the ability of two quantum conserved quantities to
serve as a CSCO set. The physical realization of our model will involve two conserved quantities in
an integrable system. Those are indeed known to realize Poisson processes \cite{bohigas1991,guhr1998},
but only after a permutation to a monotonically increasing sequence. For two observables,
these permutations are completely unrelated. Thus a physically more relevant model would
involve involve sequences that are represented by a random permutation of a Poisson sequence,
\begin{align*}
&
\left( I_{1},\,I_{2},\,\ldots,\,I_{n},\,\ldots,\,I_{N} \right) =
\\
&
\quad \mbox{Random permutation}[
      \left( \tilde{I}_{1},\,\tilde{I}_{2},\,\ldots,\,\tilde{I}_{n},\,\ldots,\,\tilde{I}_{N} \right)
                                ]
\,\,.
\end{align*}
Numerically,
those are modeled by sets of real numbers randomly independently distributed over an interval,
with {\it no subsequent reordering}.

\begin{figure}
\centering
\includegraphics[scale=.65]{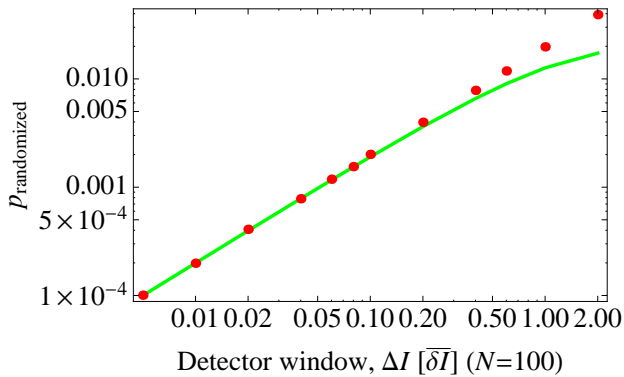}
\caption{
Probability for a given pair of consecutive elements of a Poisson distributed spectrum,
subsequently {\it randomly permuted}, to
be unresolvable by a detector with an error $\Delta I$ (i.e. to be degenerate).
$\overline{\delta I}$ is the mean spacing between consecutive elements before the random
permutation. Red dots: numerical model,
uncorrelated uniformly distributed random numbers on an interval, with {\it no further rearrangement}.
Green line: theoretical prediction (\ref{p_randomized}). Numerically,
the spectrum consists of 100 elements. Numerical points
result from an averaging over 1000 Monte Carlo realizations.  	 	
        }
\label{f:_R__ProbabilityForAPairToBeDegenerateIUNORDERED}
\end{figure}

To make a theoretical prediction for the probability for a given pair of two subsequent elements
of the sequence to be degenerate, we first assume that the probability for degenerate pairs
to share elements ({\it i.e. form clusters of degenerate pairs})
is low (see Fig.~\ref{f:_R__ProbabilityForAPairToBeLeftPairOfDegenerateClusterI}
and the expression (\ref{p_c})).
In this case, one can assume that each degenerate pair constitutes a pair of consecutive elements
in the monotonically increasing counterpart of the sequence. The probability of the later is
$2/N_{p}$ (with $N_{p}$ given by (\ref{N_p})). This probability must then be multiplied by the probability of a neighboring pair in the
monotonic sequence to be degenerate, $p$. We get
\begin{align}
\begin{aligned}
&
p_{\mbox{\scriptsize randomized}} \equiv
\\
&\qquad
\mbox{{\it Prob}}[\mbox{for a given $n$, }I_{n+1}-I_{n} < \Delta I] =
\\
&\qquad\qquad
\frac{2}{N_{p}} p
\stackrel{N\to\infty}{\approx}
\frac{2}{N} p
\,\,.
\label{p_randomized}
\end{aligned}
\end{align}
Fig.~\ref{f:_R__ProbabilityForAPairToBeDegenerateIUNORDERED} demonstrates that this
prediction is indeed correct.

\begin{figure}
\centering
\includegraphics[scale=.65]{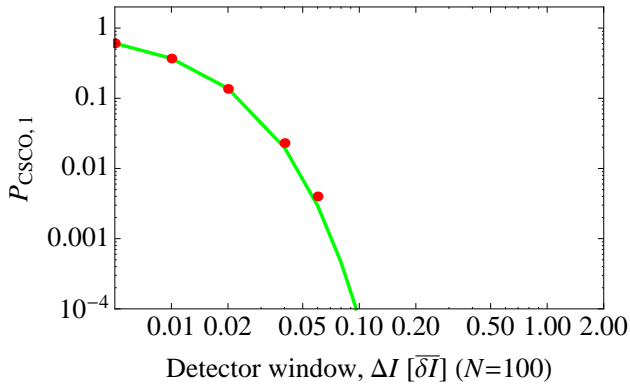}
\caption{
Probability of having no degenerate pairs of elements at all, in a Poisson distributed
spectrum (regardless of whether reshuffled or not), as a function
of the detector error $\Delta I$. Physically, this probability corresponds to the probability
for a given observable to form a complete set of commuting observables (CSCO).
Red dots: numerics. Green line: theoretical prediction (\ref{P_CSCO_1}).
The rest of
the data is the same as in Fig.~\ref{f:_R__ProbabilityForAPairToBeDegenerateI}.	
        }
\label{f:_R__ProbabilityOfNodegeneratepairsI}
\end{figure}

We are now in the position to assess the ability of a single observable $\hat{I}$ to serve as a CSCO. The
probability of that is given by the probability of having no degenerate pairs at all. Since this
probability is a property of the sequence as a set, it can be estimated using the monotonically
increasing counterpart. There, the probability of not having degenerate pairs is the probability
that none of the $N-1$ pairs of consecutive indices are degenerate. We get
\begin{align}
\begin{aligned}
&
P_{\mbox{\scriptsize CSCO, 1}} \equiv
\\
&\qquad
\mbox{{\it Prob}}[
\mbox{for no $n$, }
I_{n+1}-I_{n} < \Delta I] =
\\
&\qquad (1-p)^{N_{p}} \stackrel{p\to 0,\, N\to\infty,\, p N \to \mbox{const}}{\approx} e^{-p N}
\,\,.
\label{P_CSCO_1}
\end{aligned}
\end{align}
Fig.~\ref{f:_R__ProbabilityOfNodegeneratepairsI} shows that this probability approaches unity
only for a detector error as low as the inverse sequence length $N$. Furthermore, the expression
(\ref{P_CSCO_1}) shows that even if for a given $N$ the given observable does form a CSCO, for larger
$N$ this property dissappears.

\section{Two sequences}
We are finally ready to address the principal question posed: what is the probability that
two observables with Poisson spectra,
subsequently independently randomly permuted, $\hat{I}^{(1)}$ and $\hat{I}^{(2)}$,
constitute a complete set of commuting observables (CSCO). Mathematically,
the corresponding probability is:
\begin{align}
\begin{aligned}
&
P_{\mbox{\scriptsize CSCO, 2}} \equiv
\\
&\qquad
\mbox{{\it Prob}}[
\mbox{for {\it no} $n$, }
(I^{(1)}_{n+1}-I^{(1)}_{n} < \Delta I^{(1)})
\mbox{ and }
\\
&\qquad\qquad
(I^{(2)}_{n+1}-I^{(2)}_{n} < \Delta I^{(2)})] =
\\
&\qquad\qquad\qquad
(1-p_{\mbox{\scriptsize randomized}}^{(1)}p_{\mbox{\scriptsize randomized}}^{(2)})^{{\cal N}_{p}}
\\
&\qquad\qquad\qquad\qquad
\stackrel{N\to\infty}{\approx} e^{-2 p^{(1)}p^{(2)}}
\,\,,
\label{P_CSCO_2}
\end{aligned}
\end{align}
where $p_{\mbox{\scriptsize randomized}}^{(\alpha)} = \frac{2}{N} p^{(\alpha)}$ are the
respective probabilities for a given pair to be degenerate (see (\ref{p_randomized})),
$p^{(\alpha)}$ is the analogue for a monotonic sequence (see (\ref{p})), and
$\Delta I^{(\alpha)}$ is the respective detection error. We assume the same spacing,
$\overline{\delta I}$, for both sequences.

A naive combinatorial interpretation of this probability is the ratio between the
number of ways in which $\bar{M}^{(\alpha)} \equiv p^{(\alpha)} N_{p}$ pairs can be chosen from two $N$-element-long
sets in such a way that no two pairs coincide and its unrestricted analogue:
\begin{align}
\begin{aligned}
&
P_{\mbox{\scriptsize CSCO, 2, conjecture}} \equiv
\\
&\quad
\frac{\left({\cal N}_{p}-\bar{M}^{(1)}\right)! \left({\cal N}_{p}-\bar{M}^{(2)}\right)!}{\left({\cal N}_{p}\right)!
   \left({\cal N}_{p}-\bar{M}^{(1)}-\bar{M}^{(2)}\right)!}
\\
&\quad\quad
\stackrel{N\to\infty}{\approx} e^{-2 p^{(1)}p^{(2)}}
\,\,,
\label{P_CSCO_2_2}
\end{aligned}
\end{align}
where ${\cal N}_{p}$ is the number of pairs of $N$ objects (\ref{cal_N_p}), and $N_{p} = N-1$ is the number 
of pairs of neighboring elements in a sequence of length $N$ (see (\ref{N_p})).
In this particular case the naive model generates the correct prediction. Note
that in this case, no correlations between the degenerate pairs within a particular sequence
are involved, thus a better result.

\begin{figure}
\centering
\includegraphics[scale=.65]{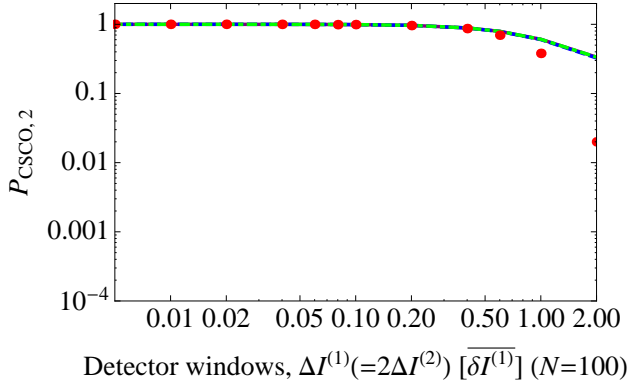}
\caption{
Probability for two Poisson-distributed spectra, subsequently independently randomly reshuffled,
not to have any degenerate pairs of elements in common. 	
Physically, this probability corresponds to the probability
for two given observables with Poisson spectra to form a complete set of commuting observables (CSCO).
Red dots: numerics. Multicolored line: comprises the theoretical prediction (\ref{P_CSCO_2}), a naive
combinatorial hypothesis
(\ref{P_CSCO_2_2}), and its large $N$ asymptotic behavior, all three mutually indistinct. Note that the former and
the latter are analytically shown to coincide.
        }
\label{f:_R__ProbabilityOfNoJointdegeneratepairsUNORDERED}
\end{figure}

\begin{figure}
\centering
\includegraphics[scale=.65]{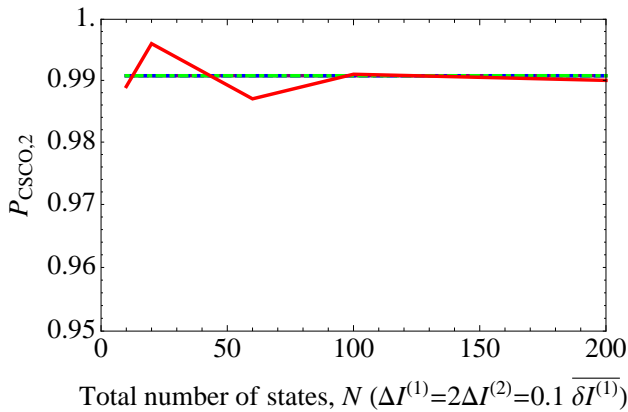}
\caption{
Numerically computed probability for two Poisson-distributed spectra, subsequently independently randomly reshuffled,
not to have any degenerate pairs of elements in common, as a function of the total length
of the spectrum $N$.
As in Fig.~\ref{f:_R__ProbabilityOfNoJointdegeneratepairsUNORDERED},
the multicolored line corresponds at the same time to the theoretical prediction (\ref{P_CSCO_2}) and to the large-$N$ asymptotic behavior
of the combinatorial expression (\ref{P_CSCO_2_2}), which are shown to coincide.
The corresponding detector window sizes are $\Delta I^{(1)}=0.1$ and $\Delta I^{(2)}=0.05$.
        }
\label{f:_R__ProbabilityOfNoJointdegeneratepairsUNORDERED_vs_N}
\end{figure}

Figure~\ref{f:_R__ProbabilityOfNoJointdegeneratepairsUNORDERED} allows one to compare the various
predictions made above, for $\Delta I^{(1)} = 2 \Delta I^{(1)} $. Notice that (a) already
at $\Delta I^{(1)} = .2$, CSCO is reached; (b) from the expression (\ref{P_CSCO_2_2}) one can see
that for large $N$, the probability (\ref{P_CSCO_2_2}) does not depend
on the length of the spectrum $N$, and thus CSCO persist for large systems. A particular
illustration of this phenomenon is presented in Fig.~\ref{f:_R__ProbabilityOfNoJointdegeneratepairsUNORDERED_vs_N}.

\section{Summary and outlook}
In this work we show that with a large probability, differing from unity only by the expression in (\ref{P_CSCO_2_2}) (see also 
Figs.~\ref{f:_R__ProbabilityOfNoJointdegeneratepairsUNORDERED}-\ref{f:_R__ProbabilityOfNoJointdegeneratepairsUNORDERED_vs_N}), two integrals 
of motion with Poisson distributed spectra form a complete set of commuting observables (CSCO); i.e. they can be used to unambiguously identify the state of the system, the detection error notwithstanding. For a given detector error, this probability converges to a fixed value as the number of states in the spectrum increases. The above result is contrasted to an analogous result for a single observable. There, (a) only for a very small detector error can this observable be used as a CSCO, and (b) for a fixed detector error, the probability of a CSCO falls to zero exponentially as the number of states in the spectrum increases. Poisson spectra constitute a popular model for energy spectra of integrable (i.e. exactly solvable) quantum systems with no true degeneracies 
\cite{bohigas1991,guhr1998}. They can also be used to model spectra of other integrals of motion of integrable systems, provided they are substantially functions of at least two quantum numbers. Note that in generic non-integrable quantum systems, energy levels repel each other 
\cite{bohigas1991,guhr1998}. There, even a poor energy detector, with an error as large as the energy level spacing, would be able to identify a state: energy can thus serve as a CSCO. Therefore, the physical implication of our principal result can be formulated as follows: given a reasonably small (i.e. smaller than the mean spacing between levels) detector error, two generic integrals of motion of an integrable quantum system can be used as a complete set of commuting observables, no matter how large the system is. On the other hand, if only one integral of motion is used, the appearance of unidentifiable states is unavoidable for large systems.

From the functional dependence perspective, we showed that in poorly measured integrable systems, all conserved quantities are (with a close-to-unity probability) functionally dependent on any two a priori chosen generic conserved quantities.

Results obtained in our article may find applications beyond the quantum state identification. They should be generally applicable in a standard pattern recognition setting where an unknown object must be identified indirectly by one, two, or more attributes. For example, using our results, we can show that chances of an ambiguity in identifying a person who was born on a particular date $D$ in a particular town $T_{1}$ within a specific time interval It and who is currently living on a particular street $S$ in another town $T_{2}$ depends neither on the length of the time interval $I_{t}$, nor on the number of the streets in $T_{2}$, but solely on the number of birth per day in $T_{1}$, probability of a further migration to $T_{2}$, percentage of $T_{1}$-born citizens in $T_{2}$, 
and the average street population in $T_{2}$; in the other words, this probability depends only on the relative measures, the absolute measures being irrelevant. The value for this probability is trivially obtainable from the central result of this paper. More generally, \emph{for a large enough set of N patterns with two Poisson distributed numerical identifiers, each measured with a finite error, the probability $P_{\mbox{\scriptsize CSCO, 2}}$ 
of having no unidentifiable patterns depends only on the probability $p^{(1)}$ of a given value of an identifier to be indistinct from its neighbor above and the analogous probability $p^{(2)}$ 
for another identifier and not on size of the set $N$}.
At the same time, if only a date of birth is known, the procedure will start frequently produce ambiguous results for long enough interval of time considered: here, the probability of finding two people born in a given town on the same date within this interval of time approaches unity. Generally, a \emph{single Poisson identifier becomes unreliable for larger sets of patterns}.


\section*{Author contributions}
Authors shared equal amount of work on the probabilistic models. E.M. provided the necessary numerical support, along with overseeing the overall 
logistics of the project. M.O. was solely responsible for the development of the combinatorial models.

\section*{Acknowledgements}
We are grateful to Vanja Dunjko and Steven G. Jackson for helpful comments and to Maxim Olshanii (Olchanyi) for providing the seed idea for this project.


\bibliographystyle{spphys}       


\end{document}